\documentclass[aps,prl,reprint,letterpaper,twocolumn,10pt,superscriptaddress]{revtex4-1}

\usepackage{graphicx}
\usepackage{amssymb, amsfonts, amsmath}
\usepackage{tikz}

\newcommand{\norm}[1]{\left\lvert{#1}\right\rvert}
\newcommand{\mom}[1]{\left\langle{#1}\right\rangle}
\newcommand{\cum}[1]{\left\langle\!\!\left\langle{#1}\right\rangle\!\!\right\rangle}

\makeatletter
\renewcommand\thefigure{\@arabic\c@figure}
\renewcommand\fnum@figure{\figurename~\thefigure}
\newcommand\colorcapon{\renewcommand\fnum@figure{\figurename~\thefigure~(color online)}}
\newcommand\colorcapoff{\renewcommand\fnum@figure{\figurename~\thefigure}}
\makeatother

\begin{document}
\title{Discrete photon statistics from continuous microwave measurements}
\author{St\'{e}phane Virally}
\email{Stephane.Virally@USherbrooke.ca}
\affiliation{D\'{e}partement de Physique, Universit\'{e} de Sherbrooke, Sherbrooke, Qu\'{e}bec J1K 2R1, Canada}
\author{Jean Olivier Simoneau}
\affiliation{D\'{e}partement de Physique, Universit\'{e} de Sherbrooke, Sherbrooke, Qu\'{e}bec J1K 2R1, Canada}
\author{Christian Lupien}
\affiliation{D\'{e}partement de Physique, Universit\'{e} de Sherbrooke, Sherbrooke, Qu\'{e}bec J1K 2R1, Canada}
\author{Bertrand Reulet}
\affiliation{D\'{e}partement de Physique, Universit\'{e} de Sherbrooke, Sherbrooke, Qu\'{e}bec J1K 2R1, Canada}
\date{\today}
\begin{abstract}
Photocount statistics are an important tool for the characterization of electromagnetic fields, especially for fields with an irrelevant phase. In the microwave domain, continuous rather than discrete measurements are the norm. Using a novel approach, we recover discrete photon stastistics from the  cumulants of a continuous distribution of field quadrature measurements. The use of cumulants allows the separation between the signal of interest and experimental noise. Using a parametric amplifier as the first stage of the amplification chain, we extract useful data from up to the sixth cumulant of the continuous distribution of a coherent field, hence recovering up to the third moment of the discrete statistics associated with a signal with much less than one average photon.
\end{abstract}

\maketitle
\emph{Introduction.} Photon statistics measurements provide a wealth of information on the state of the electromagnetic field. For instance, Glauber's theory of optical coherence~\cite{Glauber1963, *Glauber1963b} is solely based on correlations between multiple photon measurements. As the discrete nature of the interaction between light and matter is essentially a quantum feature, statistical distributions can also characterize the classicality of photonic states. For instance, single photon states~\cite{Hong1986,Eisaman2011} exhibit sub-Poisson photocount distributions that are not predicted by classical theories~\cite{Mandel1995}. These states are not just of theoretical interest, as they feature prominently in proposals for the development of quantum computation~\cite{Knill2001} and quantum communication networks~\cite{Sangouard2011}.

With the advent of circuit QED~\cite{Girvin2009}, there is currently a great interest in quantum states of the electomagnetic field in the microwave domain. From early on, predictions have been made on the specific type of photon statistics that can be expected in mesoscopic conductors~\cite{Beenakker2001,*Beenakker2004,Lebedev2010,Padurariu2012}. Recently, purely quantum photonic states have been demonstated in the microwave domain, using superconducting devices~\cite{Bozyigit2010, Lang2013}. Entanglement has also been demonstrated in the GHz range using normal conductors~\cite{Menzel2012a, Forgues2014, *Forgues2015}. The detection of single photons in the microwave domain remains a challenge, but schemes have been proposed for the extraction of photocounts from continuous measurements with linear detectors~\cite{DaSilva2010}. Experiments have already been performed to specifically extract discrete statistics from continuous measurements~\cite{Gabelli2004, Menzel2010, Zakka-Bajjani2010}, although they focused on the calculation of centered moments of the continuous distributions, rather than the cumulants, as is the case herein. 

In this Letter, we derive simple formulas linking the cumulants~\cite{Kenney1951} of the continuous variable (CV) distribution of field quadrature measurements to the centered moments of the photocount statistics. The latter fully characterize the state of the electromagnetic field when its phase is either not well defined or irrelevant. The additive nature of cumulants is especially important in the quantum regime, when we  reconstruct photon statistics for signals with much less than one photon per measurement on average and noise contributions cannot be discarded. We also use a quantum-limited parametric amplifier as the initial stage of the amplification chain, in order to lower the noise background to a few photons per measurement. With both techniques, we can characterize quantum sources with much less than one average photon in just a few seconds, without the need for excessive averaging.

This Letter is organized as follows. We first derive formulas linking the cumulants of the CV distribution to the discrete moments of the photocount distribution. We then discuss semi-classical limits and show how quantum states can be characterized from their statistics. Finally, in the experimental section, we recover the first three moments of the photocount distribution associated with a weak coherent state (featuring much less than one photon on average per measurement) by measuring the first six sumulants of the CV distribution.

\vspace{.5\baselineskip}
\emph{Theory.}
We measure quadratures of a bosonic input field by mixing the signal with a local oscillator (heterodyne interferometric measurement). If the input field is associated with the ladder operator $\hat{a}$, observables are of the form
\begin{equation}
\label{X}
\hat{X}_{\hat{a},\theta}=\frac{e^{i\theta}\,\hat{a}^\dagger+e^{-i\theta}\,\hat{a}}{\sqrt{2}},
\end{equation}
with $\theta$ the phase between the input field and the local oscillator.

We focus on the case of a measurement where the phase difference $\theta$ is effectively averaged during the full time $\tau$ of statistics accumulation. This is the case when the phase varies randomly during the detection, but also in the narrow-band limit, when the pulsations $\omega_0$ of the local oscillator and $\omega$ of the measured signal differ by $\delta\omega$, with $\delta\omega\,\tau\!\gg\!1$. As the phase $\theta$ is averaged, expectations of all odd centered moments of the measured signal are zero, while expectations of the $2k^\textrm{th}$ centered moments are
\begin{equation}
\label{X2k}
\mom{X_{\hat{a},\theta}^{2k}}_\theta=\left(\frac{1}{2}\right)^k\mom{\sum_\textrm{c.s.}\hat{a}^k\,\hat{a}^{\dagger k}},
\end{equation}
with ``c.s." standing for completely symmetric (i.e., the sum is taken on all possible permutations of the non-commuting operators). Here, $\mom{\bullet}$ represents averaging over the quantum ensemble, while $\mom{\bullet}_\theta$ represents averaging over both the quantum ensemble and the phase $\theta$. To simplify notations, we drop the $\theta$ indices in the calculations below.

The completely symmetric sum of Eq.~\ref{X2k} can be explicitely evaluated and yields~\cite{Cahill1969a}
\begin{equation}
\begin{split}
\sum_\textrm{c.s.}\hat{a}^k\,\hat{a}^{\dagger k}&=\sum_{i=0}^k\left(\frac{1}{2}\right)^{k-i}\frac{(2k)!}{(i!)^2(k-i)!}\;\hat{a}^{\dagger i}\,\hat{a}^i\\
&=\sum_{i=0}^k\left(\frac{1}{2}\right)^{k-i}\frac{(2k)!}{(i!)^2(k-i)!}\;\prod_{j=0}^{i-1}\left(\hat{n}-j\right),
\end{split}
\label{sumG}
\end{equation}
with $\hat{n}=\hat{a}^\dagger\,\hat{a}$ the usual number operator. We have explicitely linked the even moments $\mom{X_{\hat{a}}^{2k}}$ of the measured CV distribution to the moments $\mom{n^\ell}$ of the discrete photocount distribution.

As explicited before, we actually want to work with the cumulants $C_k=\cum{X_{\hat{a}}^{k}}$ of the CV distribution rather than its moments. Cumulants characterize the distribution in the exact same way as its moments, but they also have the advantage of being additive for independant distributions. This is crucial when unavoidable noise contributions, including vacuum fluctuations and the weak thermal contributions of the amplification process, are taken into account. With the use of cumulants, these independant contributions can be measured separately and removed form the final measurement, leaving only the signal, however weak.

Simple algebra leads from Eq.~\ref{sumG} to the first three moments of the photocount distribution in the form
\begin{align}
&\mom{n}=C_2-\frac{1}{2};\label{n}\\
&\mom{\delta n^2}=\frac{2}{3}\,C_4+C_2^2-\frac{1}{4};\label{n2}\\
&\mom{\delta n^3}=\frac{2}{5}\,C_6+4\,C_4\,C_2+2\,C_2^3-\frac{1}{2}\,C_2\label{n3},
\end{align}
where $\mom{\delta n^\ell}=\mom{\left(n-\mom{n}\right)^\ell}$ is the $\ell^\textrm{th}$ centered moment of the photocount distribution.

Cumulants can also be obtained through the use of the second cumulant-generating function~\cite{Kenney1951}, which is the natural logarithm of the characteristic function, or
\begin{equation}
\Gamma(\lambda)=\ln\!\left[\mom{\exp\!\left(i\lambda X_{\hat{a},\theta}\right)}_\theta\right],
\end{equation}
as
\begin{equation}
C_k=(-i)^k\;\frac{d^k\,\Gamma}{d\,\lambda^k}(0).
\end{equation}

The second cumulant-generating function can be computed for some usual states~\cite{Gardiner2004}. For a thermal state, $\Gamma(\lambda)=-(\mom{n}+1/2)\,\lambda^2/2$. This is the second cumulant-generating function of a gaussian distribution, leading to $C_2=\mom{n}+1/2$ and $C_k=0,\;\forall k\ge3$. In other words, there is an equivalence between gaussian distribution of quadratures and thermal (chaotic) statistics. When the fourth-order cumulant is not zero, the statistics is not chaotic. This is in contradiction with a claim made in ref.~\cite{Zakka-Bajjani2010} for the noise emitted by a tunnel junction. Such noise is known to be caused by Poissonian charge tunneling statistics~\cite{Blanter2000} and thus features a non-vanishing fourth-order current cumulant ($C_4\propto e^3\,B^3\,\bar{I}$ where $e$ is the unit charge, $B$ the bandwidth and $\bar{I}$ the average current). For this reason, the associated photocount statistics is not chaotic. However, the ratio $C_4/C_2^2\propto e\,B/\bar{I}$ of the current statistics is typically small ($10^{-4}$ to $10^{-3}$ in~\cite{Zakka-Bajjani2010}). It can then be difficult to discern between chaotic and non-chaotic statistics, especially in the presence of detection noise.

For a coherent state, $\Gamma(\lambda)=-\lambda^2/4+\ln\!\left[I_0(i\sqrt{2}\norm{\alpha}\lambda)\right]$, where $I_0$ is the modified Bessel function of the first kind. This leads to $C_2=\norm{\alpha}^2+1/2$, $C_4=-3\norm{\alpha}^4/2$ and $C_6=10\norm{\alpha}^6$. Using Eqs.~\ref{n},~\ref{n2},~\ref{n3} we get, as expected, $\mom{n}=\mom{\delta n^2}=\mom{\delta n^3}=\norm{\alpha}^2$.

\vspace{.5\baselineskip}
\emph{Semi-classical narrow-band model.}
We wish to extract useful limits for semi-classical narrow-band signals. We first model the amplitude of a fully classical narrow-band signal as
\begin{equation}
s(t)=a(t)\cos(\omega t)+b(t)\sin(\omega t),
\end{equation}
where $a$ and $b$ are real functions of time with slow variations (in the sense that the time scale of their variations is much smaller than $1/\omega$). We consider a measurement process that cannot resolve the fast oscillations at $\omega$. The intensity of a such a signal is then defined by
\begin{equation}
i(t)=\frac{1}{2}\left[a^2(t)+b^2(t)\right].
\end{equation}
A semi-classical prolongation of the model requires adding an ersatz for the vacuum, in the form 
\begin{equation}
s_\mathrm{vac}(t)=\xi_a(t)\cos(\omega t)+\xi_b(t)\sin(\omega t),
\end{equation}where $\xi_a$ and $\xi_b$ are two independant gaussian random processes with null average and $1/2$ variance.

By identifying $X_{\hat{a},\omega t}$ with $s(t)+s_\mathrm{vac}(t)$, and applying Eqs.~\ref{n},~\ref{n2},~\ref{n3} we get (after averaging $\omega t$ over many cycles)
\begin{align}
&\mom{n}=\mom{i};\label{momI}\\
&\mom{\delta n^2}=\mom{i}+\mom{\delta i^2},\label{deltaI2}\\
&\mom{\delta n^3}=\mom{i}+3\mom{\delta i^2}+\mom{\delta i^3},\label{deltaI3}
\end{align}
where $\mom{\delta i^k}$ is the $k^\mathrm{th}$ centered moment of the intensity distribution.

Eqs.~\ref{deltaI2,deltaI3}  provide interesting limits for classical signals. There exists inequalities verified by continuous distributions. For instance$\mom{\delta i^2}\ge0$ implies
\begin{equation}
\label{dn2lim}
\mom{\delta n^2}\ge\mom{n}.
\end{equation}
Thus, we recover the fact that classical signals cannot obey sub-poissonian statistics~\cite{Loudon2000}. When the intensity is constant, all moments $\mom{\delta i^k}$ vanish and we recognize the Poisson statistics of coherent signals. Super-poissonian statistics exists as soon as $i(t)$ is not constant. This is the case, for instance, when the intensity is modulated, as shown in the experimental section.

Additional inequalities stem from Stieltjes moments theorem~\cite{Widder1946}. For instance, $\mom{\delta i^3}\ge-3\mom{i}\mom{\delta i^2}$ leads to
\begin{equation}
\label{dn3lim}
\mom{\delta n^3}\ge\mom{n}+3\left(\mom{\delta n^2}-\mom{n}\right)\left(1-\mom{n}\right).
\end{equation}
In particular, if $\mom{n}\le1$, $\mom{\delta n^3}\ge\mom{n}$.

In addition to the ``hard'' inequalities of Eqs.~\ref{dn2lim},~\ref{dn3lim}, we can derive relations in the vanishing signal limit. For a classical signal, a reduction in intensity is equivalent to the addition of an attenuator in the line. It is possible to write $\mom{i}=\eta\mom{I}$ and $\mom{\delta i^k}=\eta^k\mom{\delta I^k}$ with a fixed intensity $I$ and a varying attenuation factor $\eta$. Accordingly, the vanishing signal limit $\mom{i}\rightarrow0$ is equivalent to the limit $\eta\rightarrow0$. Thus, Eqs.~\ref{momI},~\ref{deltaI2},~\ref{deltaI3} yield $\mom{\delta n^2}\!,\,\mom{\delta n^3}\sim\mom{n}$ as $\mom{n}\rightarrow0$. Hence, the Fano factor~\footnote{Alternatively, it is common in quantum optical experiments to measure related quantities such as Mandel's $Q$ parameter ($Q=\protect\mathcal{F}-1$) or the degree of second-order coherence ($g^{(2)}(0)=1+Q/\protect\langle n\protect\rangle$).} $\mathcal{F}=\mom{\delta n^2}/\mom{n}$ of a narrow-band classical signal verifies
\begin{equation}
\label{Fano}
\mathcal{F}\rightarrow1\textrm{ as }\mom{n}\rightarrow0.
\end{equation}

In fact, the Fano factor is a great tool for characterizing distributions in the low $\mom{n}$ limit. It should always be close to 1 for classical states and only deviate reasonably from this value for non-classical states such as Fock states or squeezed vacua. Although Fano factors of a few hundreds have been reported in previous experiments~\cite{Gabelli2004}, they only reflect the fact that moments were used, rather than cumulants, thus mixing signal and noise. They do not reflect the underlying statistics of coherent states.

\begin{figure}[htb]
\centering
\includegraphics[width=.5\textwidth]{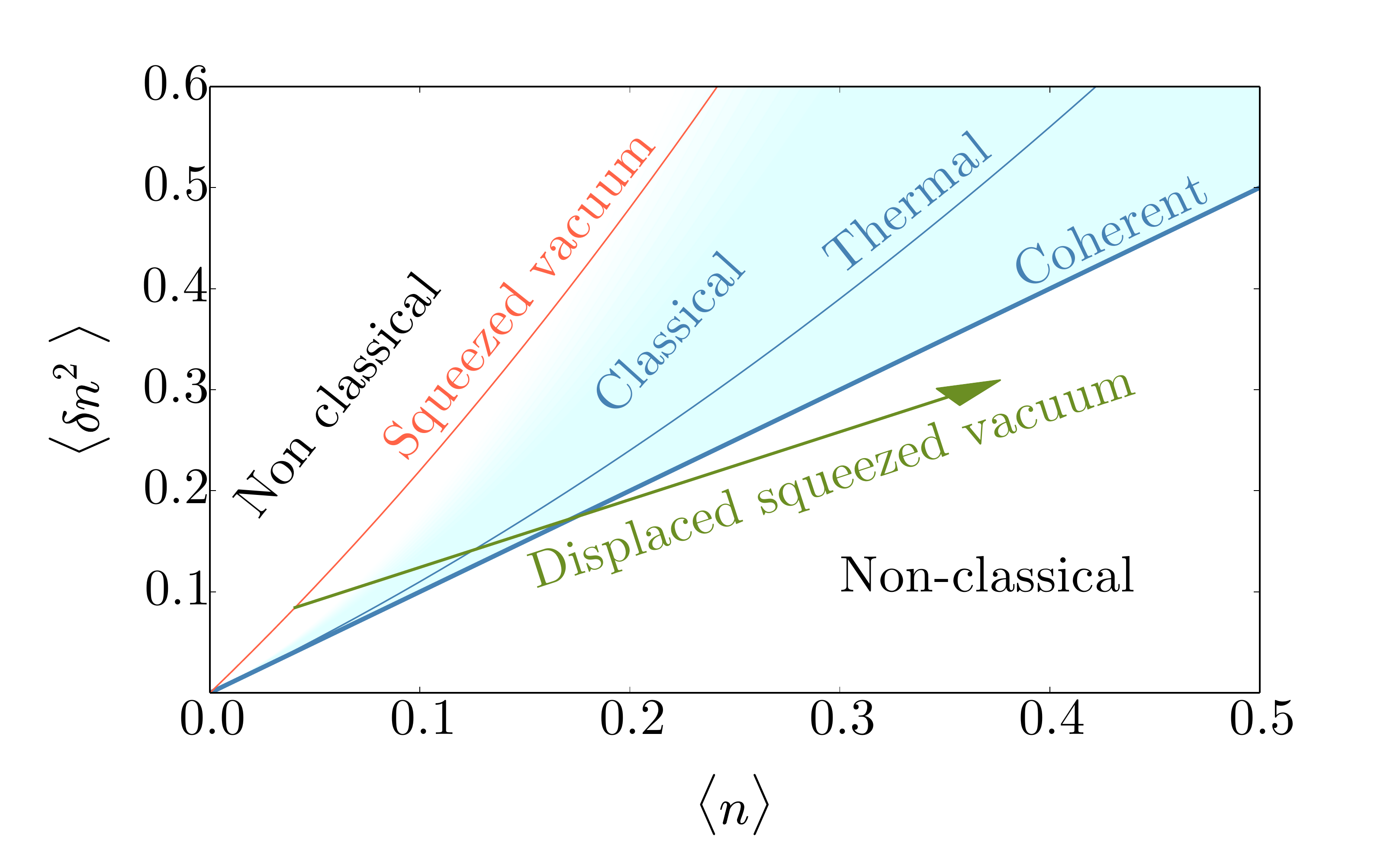}
\colorcapon
\caption{States of a narrow-band semi-classical electromagnetic field are restricted in the $(\mom{n},\mom{\delta n^2})$ plane. They can only exist in the shaded region. The lower limit (thick blue line) represents coherent states and poissonian statistics. The shaded region is not bounded by a ``hard'' upper curve as classical states are only constrained in the limit $\mom{n}\rightarrow0$. Squeezed vacua (solid red line) can exist in the non-classical upper part. They can cross from the upper to the lower non-classical parts when appropriately displaced (green arrow).}
\colorcapoff
\label{CNC}
\end{figure}

Fig.~\ref{CNC} identifies classical and non-classical regions in the $\left(\mom{n},\mom{\delta n^2}\right)$ plane. For a given photon average, coherent states feature the lowest variance of all classical states. As shown above, there also exists an upper limit for narrow-band classial signals as the photon average vanishes, with a Fano factor approaching unity. There are thus two distinct non-classical zones in the plane. Quantum states below the $\mom{\delta n^2}=\mom{n}$ line (i.e. states with sub-Poisson statistics) are well-known and include Fock states (on the x-axis). On the other side, squeezed vacuum states exhibit super-poissonian statistics even as $\mom{n}\rightarrow0$ (with a limit $\mom{\delta n^2}\sim2\mom{n}$ that reflects the production of photon pairs in these states). Although the two non-classical regions seem segregated, it is possible in some cases to move a state from one to the other using a symplectic operation, such as a displacement. For instance, when a squeezed vacuum is displaced in the quadrature plane in the direction of squeezing, it can cross the clasical region and end up as an amplitude squeezed state in the sub-Poisson region~\cite{Grosse2007, Lemonde2014}. Indeed, under application of the Glauber displacement operator~\cite{Loudon2000} $\hat{D}\!\left(r\,e^{i\theta}\right)$, the average photon number and variance of the photocount distribution become
\begin{equation}
\label{nD}
\mom{n}\rightarrow\mom{n}+\sqrt{2}\,r\mom{X_{\hat{a},\theta}}+r^2
\end{equation}
and
\begin{equation}
\label{dn2D}
\mom{\delta n^2}\rightarrow\mom{\delta n^2}+2\,r^2\mom{\delta\hat{X}_{\hat{a},\theta}^2}+2\sqrt{2}\,r\,\textrm{cov}(\hat{n},\hat{X}_{\hat{a},\theta}),
\end{equation}
where the covariance between two operators is defined as
\begin{equation}
\textrm{cov}\!\left(\hat{p},\hat{q}\right)=\frac{1}{2}\left(\mom{\hat{p}\,\hat{q}}+\mom{\hat{q}\,\hat{p}}\right)-\mom{\hat{p}}\mom{\hat{q}}.
\end{equation}

The migration from one non-classical region to the other is due to the presence of the covariance term that reflects correlations between $\hat{n}$ and $\hat{X}_{\hat{a},\theta}$.

\begin{figure}[htb]
\centering
\includegraphics[width=.47\textwidth]{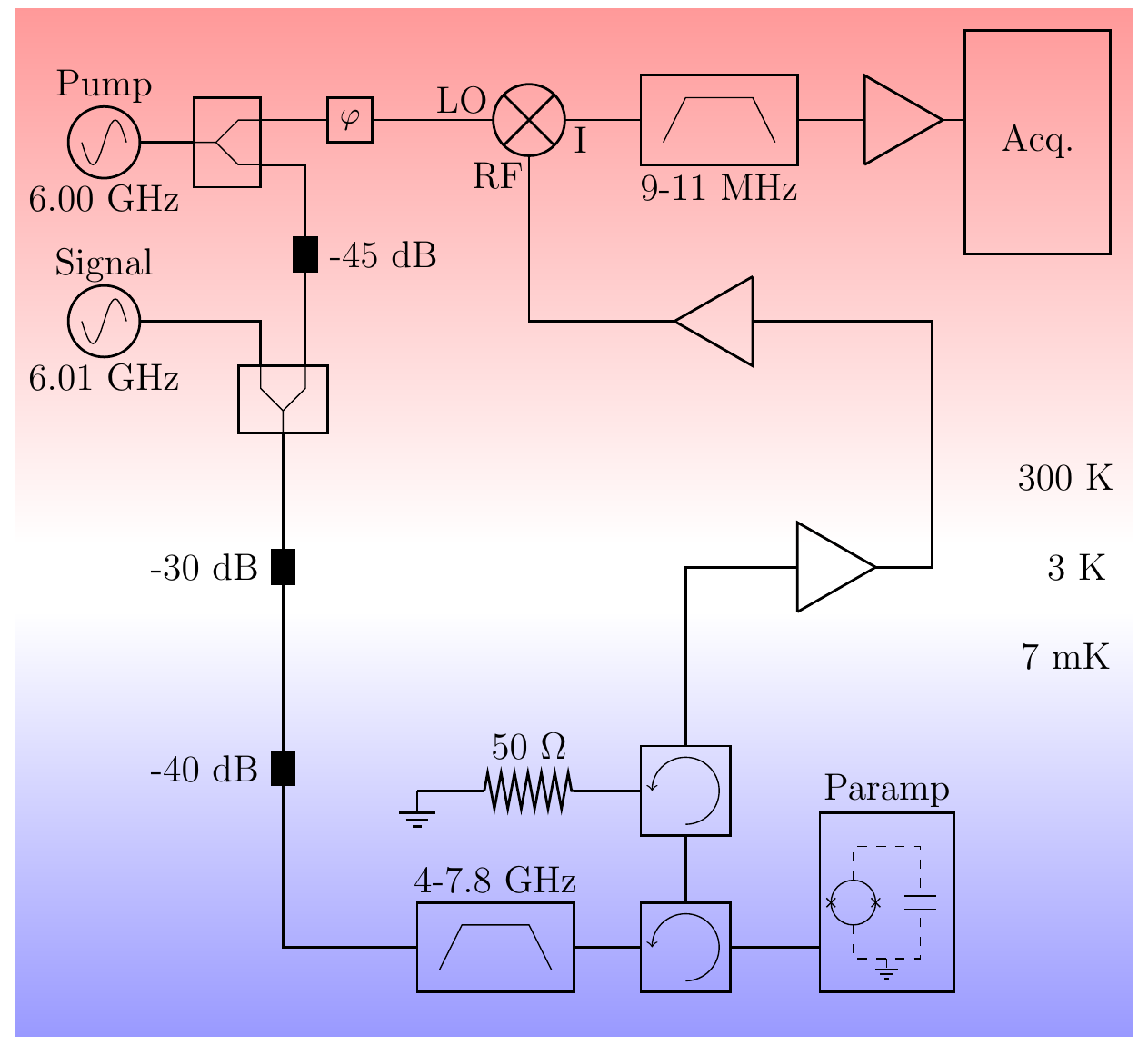}
\colorcapon
\caption{Experimental setup (details in text).}
\colorcapoff
\label{exp}
\end{figure}

\vspace{.5\baselineskip}
\emph{Experiment.}
We tested the results of Eqs.~\ref{n2},~\ref{n3} on a weak coherent state. The full experimental setup is represented in Fig.~\ref{exp}. The ideal signal is well approximated by a phase and amplitude-controlled sine wave at 6.01~GHz generated by an analog signal generator. It is attenuated down to much less than a photon per measurement bin before going through a Josephson parametric amplifier (paramp)~\cite{Hatridge2011} placed in a dilution refrigerator  and thermalized at 7~mK. The parametric amplifier is pumped in voltage by a second signal from an analog signal generator at 6.00~GHz. The measured gain is close to 15~dB. The parametric amplifier is protected from the noise of the rest of the amplification chain by two cryogenic circulators. The amplification chain continues with secondary cryogenic low noise amplifier thermalized at 3~K, and additional amplifiers at room temperature. The parametric amplifier pump is also used as a local oscillator for an in-phase/quadrature mixer. The phase of the local oscillator is controlled by a phase shifter. The signal at 6.01~GHz is thus downconverted to 10~MHz, band-pass filtered and acquired by a fast acquisition card with 14-bit resolution and a 400~MSa/s sampling rate.

Cumulants are extracted from raw power data using a model with a single free parameter (number of noise of photons in the cavity of the parametric amplifier). The value of this parameter is obtained by replacing the signal source by a DC-biased tunnel junction (not shown) placed in the dilution refrigerator. The characteristic noise curve is fitted to extract the background noise, about one average photon in the 50-MHz bandwidth cavity of the parametric amplifier (-125~dBm). Experimental data is obtained by sweeping the signal source power in order to get from $10^{-3}$ signal photon on average (-155~dBm) to 10 signal photons on average (-115~dBm) in the cavity. Data above one average signal photon was discarded due to the very nonlinear response of the parametric amplifier.

\begin{figure}[htb]
\centering
\includegraphics[width=.5\textwidth]{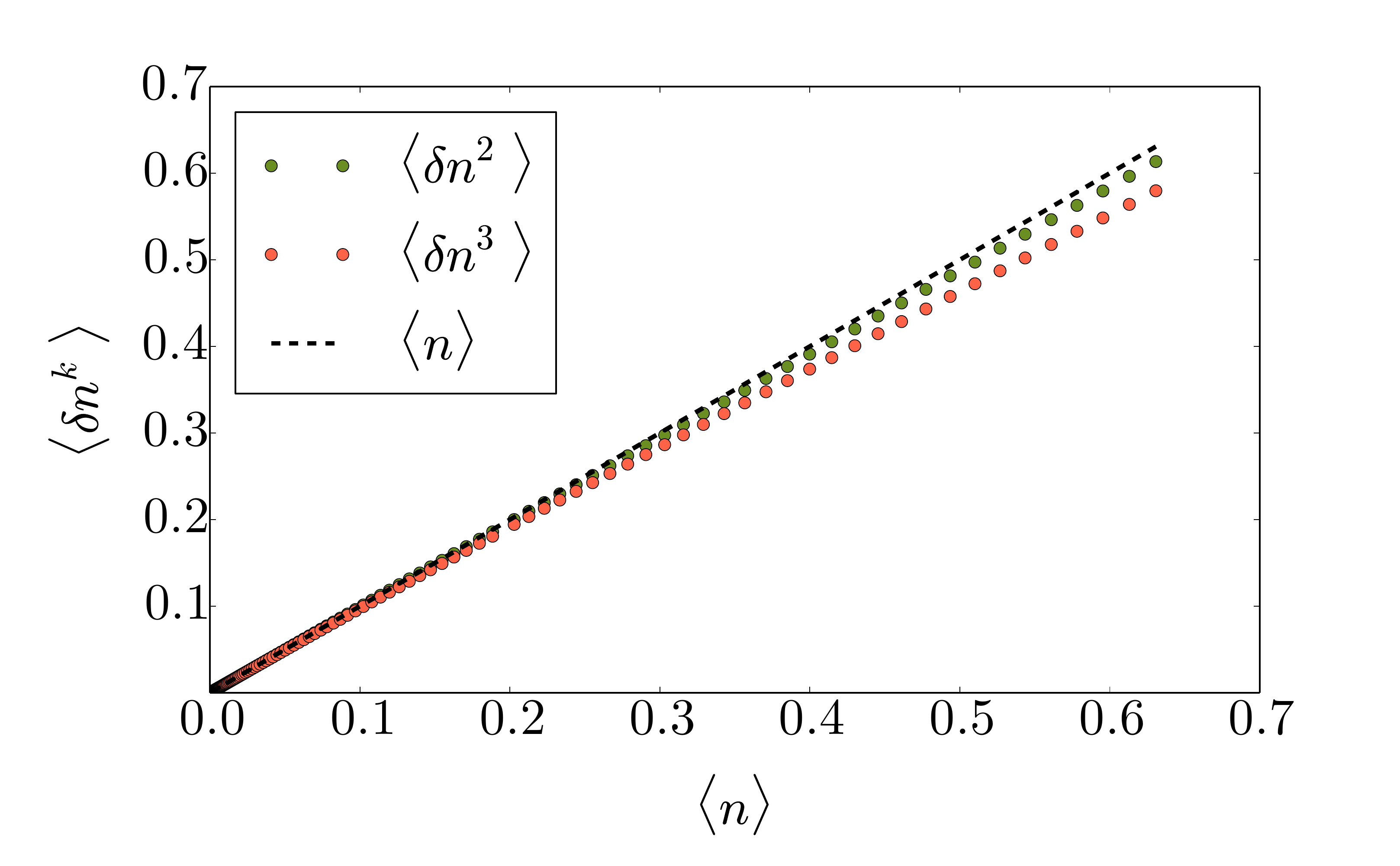}
\colorcapon
\caption{Measured variance and skewness of photon statistics for a coherent state. All centered moments of the Poisson distribution verify $\mom{\delta n^k}=\mom{n}$.}
\colorcapoff
\label{dnk}
\end{figure}

The first six cumulants of the CV distribution are computed in real time, allowing the extraction of the first three moments of the photocount distribution. Results are presented in Fig.~\ref{dnk}. Each point of the graph is computed from close to 800~million samples and is acquired and processed in about 2~s. Experimental results agree remarkably well with expected values for a coherent state, for which the Poisson distribution verifies $\mom{\delta n^3}=\mom{\delta n^2}=\mom{n}$. The small deviation at higher $\mom{n}$ can be ascribed to the strong nonlinearity of the parametric amplifier. Essential features in the graph, including the unit Fano factor, are robust with respect to variations in the value of the free parameter.

\begin{figure}[htb]
\centering
\includegraphics[width=.5\textwidth]{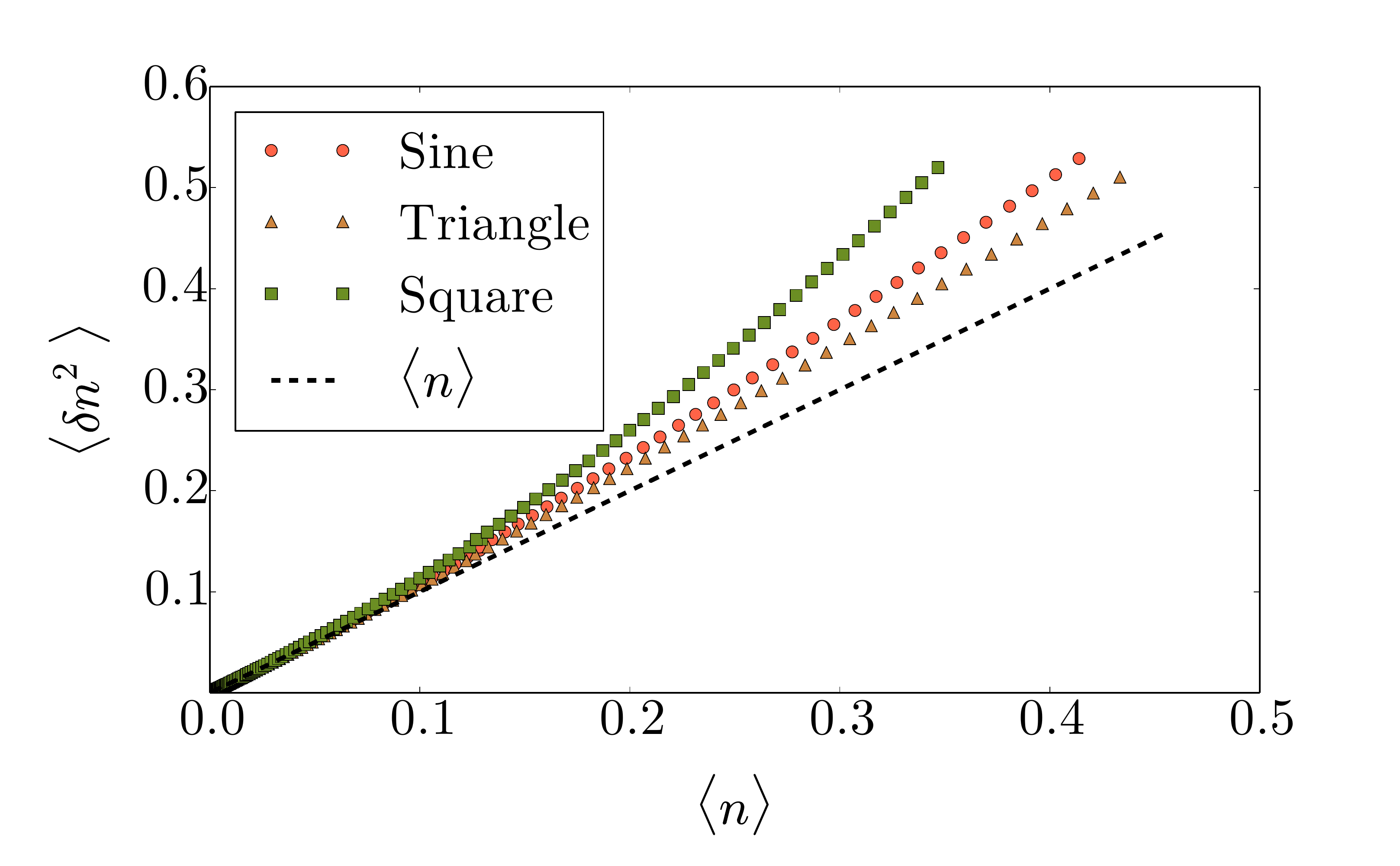}
\colorcapon
\caption{Variance of photon statistics for different modulations schemes. Although all distributions are super-poissonian, in the small signal limit all variances move towards $\mom{n}$, as expected for classical states.}
\colorcapoff
\label{Mod}
\end{figure}

Fig.~\ref{Mod} illustrates the positive contribution of intensity fluctuations in the variance of the measured signal. This set of data was obtained by slowly modulating the coherent state of Fig.~\ref{dnk} with various schemes. The modulation rate was slightly less than 100~kHz, at a rate incommensurable with the frequency of the coherent state and that of the pump/local oscillator. The various modulation schemes correspond to different variances of the intensity distribution. All variances are of the form $\mom{\delta i^2}=\alpha\mom{i}^2$, with $\alpha=1/3$ for triangular modulation, $\alpha=1/2$ for sinusoidal modulation, and $\alpha=1$ for square modulation. The unit Fano factors predicted by Eq.~\ref{Fano} are clearly visible in Fig.~\ref{Mod} (which can be compared to Fig.~\ref{CNC}).

\vspace{.5\baselineskip}
\emph{Conclusion.}
Discrete photon statistics can be recovered from the cumulants of the continuous distribution of field quadratures. Cumulants allow the signal to be completely separated from the noise, leading to measurements of signals with much less than one average photon. We performed a proof of principle experiment with coherent states and modulated signals featuring super-poissonian statistics. Thanks to the segregation between signal and noise, and the use of a parametric amplifier, we recovered the first three moments of the photocount statistics. This type of experiment can differentiate between classical and non-classical states of the microwave radiation in mesoscopic conductors. In particular, the sub-poissonian statistics of single photon sources would be recognized. In the case of squeezed vacua, the generation of pairs of photons would yield a non-classical Fano factor of 2 in the limit of vanishing $\mom{n}$.

\vspace{\baselineskip}
We acknowledge fruitful discussions with J.~Gabelli. We thank G.~Lalibert\'{e} for technical help. This work was supported by the Canada Excellence Research Chairs program, Canada NSERC, Qu\'{e}bec MEIE, Qu\'{e}bec FRQNT via INTRIQ, Universit\'{e} de Sherbrooke via EPIQ, and Canada Foundation for Innovation.

\bibliographystyle{apsrev4-1}
\bibliography{Pstats}

\begin{thebibliography}{31}%
\makeatletter
\providecommand \@ifxundefined [1]{%
 \@ifx{#1\undefined}
}%
\providecommand \@ifnum [1]{%
 \ifnum #1\expandafter \@firstoftwo
 \else \expandafter \@secondoftwo
 \fi
}%
\providecommand \@ifx [1]{%
 \ifx #1\expandafter \@firstoftwo
 \else \expandafter \@secondoftwo
 \fi
}%
\providecommand \natexlab [1]{#1}%
\providecommand \enquote  [1]{``#1''}%
\providecommand \bibnamefont  [1]{#1}%
\providecommand \bibfnamefont [1]{#1}%
\providecommand \citenamefont [1]{#1}%
\providecommand \href@noop [0]{\@secondoftwo}%
\providecommand \href [0]{\begingroup \@sanitize@url \@href}%
\providecommand \@href[1]{\@@startlink{#1}\@@href}%
\providecommand \@@href[1]{\endgroup#1\@@endlink}%
\providecommand \@sanitize@url [0]{\catcode `\\12\catcode `\$12\catcode
  `\&12\catcode `\#12\catcode `\^12\catcode `\_12\catcode `\%12\relax}%
\providecommand \@@startlink[1]{}%
\providecommand \@@endlink[0]{}%
\providecommand \url  [0]{\begingroup\@sanitize@url \@url }%
\providecommand \@url [1]{\endgroup\@href {#1}{\urlprefix }}%
\providecommand \urlprefix  [0]{URL }%
\providecommand \Eprint [0]{\href }%
\providecommand \doibase [0]{http://dx.doi.org/}%
\providecommand \selectlanguage [0]{\@gobble}%
\providecommand \bibinfo  [0]{\@secondoftwo}%
\providecommand \bibfield  [0]{\@secondoftwo}%
\providecommand \translation [1]{[#1]}%
\providecommand \BibitemOpen [0]{}%
\providecommand \bibitemStop [0]{}%
\providecommand \bibitemNoStop [0]{.\EOS\space}%
\providecommand \EOS [0]{\spacefactor3000\relax}%
\providecommand \BibitemShut  [1]{\csname bibitem#1\endcsname}%
\let\auto@bib@innerbib\@empty
\bibitem [{\citenamefont {Glauber}(1963{\natexlab{a}})}]{Glauber1963}%
  \BibitemOpen
  \bibfield  {author} {\bibinfo {author} {\bibfnamefont {R.~J.}\ \bibnamefont
  {Glauber}},\ }\href@noop {} {\bibfield  {journal} {\bibinfo  {journal}
  {Physical Review}\ }\textbf {\bibinfo {volume} {130}},\ \bibinfo {pages}
  {2529} (\bibinfo {year} {1963}{\natexlab{a}})}\BibitemShut {NoStop}%
\bibitem [{\citenamefont {Glauber}(1963{\natexlab{b}})}]{Glauber1963b}%
  \BibitemOpen
  \bibfield  {author} {\bibinfo {author} {\bibfnamefont {R.~J.}\ \bibnamefont
  {Glauber}},\ }\href@noop {} {\bibfield  {journal} {\bibinfo  {journal}
  {Physical Review Letters}\ }\textbf {\bibinfo {volume} {10}},\ \bibinfo
  {pages} {84} (\bibinfo {year} {1963}{\natexlab{b}})}\BibitemShut {NoStop}%
\bibitem [{\citenamefont {Hong}\ and\ \citenamefont {Mandel}(1986)}]{Hong1986}%
  \BibitemOpen
  \bibfield  {author} {\bibinfo {author} {\bibfnamefont {C.~K.}\ \bibnamefont
  {Hong}}\ and\ \bibinfo {author} {\bibfnamefont {L.}~\bibnamefont {Mandel}},\
  }\href@noop {} {\bibfield  {journal} {\bibinfo  {journal} {Physical Review
  Letters}\ }\textbf {\bibinfo {volume} {56}},\ \bibinfo {pages} {58} (\bibinfo
  {year} {1986})}\BibitemShut {NoStop}%
\bibitem [{\citenamefont {Eisaman}\ \emph {et~al.}(2011)\citenamefont
  {Eisaman}, \citenamefont {Fan}, \citenamefont {Migdall},\ and\ \citenamefont
  {Polyakov}}]{Eisaman2011}%
  \BibitemOpen
  \bibfield  {author} {\bibinfo {author} {\bibfnamefont {M.}~\bibnamefont
  {Eisaman}}, \bibinfo {author} {\bibfnamefont {J.}~\bibnamefont {Fan}},
  \bibinfo {author} {\bibfnamefont {A.}~\bibnamefont {Migdall}}, \ and\
  \bibinfo {author} {\bibfnamefont {S.}~\bibnamefont {Polyakov}},\ }\href@noop
  {} {\bibfield  {journal} {\bibinfo  {journal} {Review of Scientific
  Instruments}\ }\textbf {\bibinfo {volume} {82}},\ \bibinfo {pages} {071101}
  (\bibinfo {year} {2011})}\BibitemShut {NoStop}%
\bibitem [{\citenamefont {Mandel}\ and\ \citenamefont
  {Wolf}(1995)}]{Mandel1995}%
  \BibitemOpen
  \bibfield  {author} {\bibinfo {author} {\bibfnamefont {L.}~\bibnamefont
  {Mandel}}\ and\ \bibinfo {author} {\bibfnamefont {E.}~\bibnamefont {Wolf}},\
  }\href@noop {} {\emph {\bibinfo {title} {{Optical Coherence and Quantum
  Optics}}}}\ (\bibinfo  {publisher} {Cambridge University Press},\ \bibinfo
  {year} {1995})\BibitemShut {NoStop}%
\bibitem [{\citenamefont {Knill}\ \emph {et~al.}(2001)\citenamefont {Knill},
  \citenamefont {Laflamme},\ and\ \citenamefont {Milburn}}]{Knill2001}%
  \BibitemOpen
  \bibfield  {author} {\bibinfo {author} {\bibfnamefont {E.}~\bibnamefont
  {Knill}}, \bibinfo {author} {\bibfnamefont {R.}~\bibnamefont {Laflamme}}, \
  and\ \bibinfo {author} {\bibfnamefont {G.~J.}\ \bibnamefont {Milburn}},\
  }\href@noop {} {\bibfield  {journal} {\bibinfo  {journal} {Nature}\ }\textbf
  {\bibinfo {volume} {409}},\ \bibinfo {pages} {46} (\bibinfo {year}
  {2001})}\BibitemShut {NoStop}%
\bibitem [{\citenamefont {Sangouard}\ \emph {et~al.}(2011)\citenamefont
  {Sangouard}, \citenamefont {Simon}, \citenamefont {{De Riedmatten}},\ and\
  \citenamefont {Gisin}}]{Sangouard2011}%
  \BibitemOpen
  \bibfield  {author} {\bibinfo {author} {\bibfnamefont {N.}~\bibnamefont
  {Sangouard}}, \bibinfo {author} {\bibfnamefont {C.}~\bibnamefont {Simon}},
  \bibinfo {author} {\bibfnamefont {H.}~\bibnamefont {{De Riedmatten}}}, \ and\
  \bibinfo {author} {\bibfnamefont {N.}~\bibnamefont {Gisin}},\ }\href@noop {}
  {\bibfield  {journal} {\bibinfo  {journal} {Reviews of Modern Physics}\
  }\textbf {\bibinfo {volume} {83}},\ \bibinfo {pages} {33} (\bibinfo {year}
  {2011})}\BibitemShut {NoStop}%
\bibitem [{\citenamefont {Girvin}\ \emph {et~al.}(2009)\citenamefont {Girvin},
  \citenamefont {Devoret},\ and\ \citenamefont {Schoelkopf}}]{Girvin2009}%
  \BibitemOpen
  \bibfield  {author} {\bibinfo {author} {\bibfnamefont {S.~M.}\ \bibnamefont
  {Girvin}}, \bibinfo {author} {\bibfnamefont {M.~H.}\ \bibnamefont {Devoret}},
  \ and\ \bibinfo {author} {\bibfnamefont {R.~J.}\ \bibnamefont {Schoelkopf}},\
  }\href@noop {} {\bibfield  {journal} {\bibinfo  {journal} {Physica Scripta}\
  }\textbf {\bibinfo {volume} {T137}},\ \bibinfo {pages} {014012} (\bibinfo
  {year} {2009})}\BibitemShut {NoStop}%
\bibitem [{\citenamefont {Beenakker}\ and\ \citenamefont
  {Schomerus}(2001)}]{Beenakker2001}%
  \BibitemOpen
  \bibfield  {author} {\bibinfo {author} {\bibfnamefont {C.~W.~J.}\
  \bibnamefont {Beenakker}}\ and\ \bibinfo {author} {\bibfnamefont
  {H.}~\bibnamefont {Schomerus}},\ }\href@noop {} {\bibfield  {journal}
  {\bibinfo  {journal} {Physical Review Letters}\ }\textbf {\bibinfo {volume}
  {86}},\ \bibinfo {pages} {700} (\bibinfo {year} {2001})}\BibitemShut
  {NoStop}%
\bibitem [{\citenamefont {Beenakker}\ and\ \citenamefont
  {Schomerus}(2004)}]{Beenakker2004}%
  \BibitemOpen
  \bibfield  {author} {\bibinfo {author} {\bibfnamefont {C.~W.~J.}\
  \bibnamefont {Beenakker}}\ and\ \bibinfo {author} {\bibfnamefont
  {H.}~\bibnamefont {Schomerus}},\ }\href@noop {} {\bibfield  {journal}
  {\bibinfo  {journal} {Physical Review Letters}\ }\textbf {\bibinfo {volume}
  {93}},\ \bibinfo {pages} {096801} (\bibinfo {year} {2004})}\BibitemShut
  {NoStop}%
\bibitem [{\citenamefont {Lebedev}\ \emph {et~al.}(2010)\citenamefont
  {Lebedev}, \citenamefont {Lesovik},\ and\ \citenamefont
  {Blatter}}]{Lebedev2010}%
  \BibitemOpen
  \bibfield  {author} {\bibinfo {author} {\bibfnamefont {A.~V.}\ \bibnamefont
  {Lebedev}}, \bibinfo {author} {\bibfnamefont {G.~B.}\ \bibnamefont
  {Lesovik}}, \ and\ \bibinfo {author} {\bibfnamefont {G.}~\bibnamefont
  {Blatter}},\ }\href@noop {} {\bibfield  {journal} {\bibinfo  {journal}
  {Physical Review B}\ }\textbf {\bibinfo {volume} {81}},\ \bibinfo {pages}
  {155421} (\bibinfo {year} {2010})}\BibitemShut {NoStop}%
\bibitem [{\citenamefont {Padurariu}\ \emph {et~al.}(2012)\citenamefont
  {Padurariu}, \citenamefont {Hassler},\ and\ \citenamefont
  {Nazarov}}]{Padurariu2012}%
  \BibitemOpen
  \bibfield  {author} {\bibinfo {author} {\bibfnamefont {C.}~\bibnamefont
  {Padurariu}}, \bibinfo {author} {\bibfnamefont {F.}~\bibnamefont {Hassler}},
  \ and\ \bibinfo {author} {\bibfnamefont {Y.~V.}\ \bibnamefont {Nazarov}},\
  }\href@noop {} {\bibfield  {journal} {\bibinfo  {journal} {Physical Review
  B}\ }\textbf {\bibinfo {volume} {86}},\ \bibinfo {pages} {054514} (\bibinfo
  {year} {2012})}\BibitemShut {NoStop}%
\bibitem [{\citenamefont {Bozyigit}\ \emph {et~al.}(2011)\citenamefont
  {Bozyigit}, \citenamefont {Lang}, \citenamefont {Steffen}, \citenamefont
  {Fink}, \citenamefont {Baur}, \citenamefont {Bianchetti}, \citenamefont
  {Leek}, \citenamefont {Filipp}, \citenamefont {da~Silva}, \citenamefont
  {Blais},\ and\ \citenamefont {Wallraff}}]{Bozyigit2010}%
  \BibitemOpen
  \bibfield  {author} {\bibinfo {author} {\bibfnamefont {D.}~\bibnamefont
  {Bozyigit}}, \bibinfo {author} {\bibfnamefont {C.}~\bibnamefont {Lang}},
  \bibinfo {author} {\bibfnamefont {L.}~\bibnamefont {Steffen}}, \bibinfo
  {author} {\bibfnamefont {J.~M.}\ \bibnamefont {Fink}}, \bibinfo {author}
  {\bibfnamefont {M.}~\bibnamefont {Baur}}, \bibinfo {author} {\bibfnamefont
  {R.}~\bibnamefont {Bianchetti}}, \bibinfo {author} {\bibfnamefont {P.~J.}\
  \bibnamefont {Leek}}, \bibinfo {author} {\bibfnamefont {S.}~\bibnamefont
  {Filipp}}, \bibinfo {author} {\bibfnamefont {M.~P.}\ \bibnamefont
  {da~Silva}}, \bibinfo {author} {\bibfnamefont {A.}~\bibnamefont {Blais}}, \
  and\ \bibinfo {author} {\bibfnamefont {A.}~\bibnamefont {Wallraff}},\
  }\href@noop {} {\bibfield  {journal} {\bibinfo  {journal} {Nature Physics}\
  }\textbf {\bibinfo {volume} {7}},\ \bibinfo {pages} {154} (\bibinfo {year}
  {2011})}\BibitemShut {NoStop}%
\bibitem [{\citenamefont {Lang}\ \emph {et~al.}(2013)\citenamefont {Lang},
  \citenamefont {Eichler}, \citenamefont {Steffen}, \citenamefont {Fink},
  \citenamefont {Woolley}, \citenamefont {Blais},\ and\ \citenamefont
  {Wallraff}}]{Lang2013}%
  \BibitemOpen
  \bibfield  {author} {\bibinfo {author} {\bibfnamefont {C.}~\bibnamefont
  {Lang}}, \bibinfo {author} {\bibfnamefont {C.}~\bibnamefont {Eichler}},
  \bibinfo {author} {\bibfnamefont {L.}~\bibnamefont {Steffen}}, \bibinfo
  {author} {\bibfnamefont {J.~M.}\ \bibnamefont {Fink}}, \bibinfo {author}
  {\bibfnamefont {M.~J.}\ \bibnamefont {Woolley}}, \bibinfo {author}
  {\bibfnamefont {A.}~\bibnamefont {Blais}}, \ and\ \bibinfo {author}
  {\bibfnamefont {A.}~\bibnamefont {Wallraff}},\ }\href@noop {} {\bibfield
  {journal} {\bibinfo  {journal} {Nature Physics}\ }\textbf {\bibinfo {volume}
  {9}},\ \bibinfo {pages} {345} (\bibinfo {year} {2013})}\BibitemShut {NoStop}%
\bibitem [{\citenamefont {Menzel}\ \emph {et~al.}(2012)\citenamefont {Menzel},
  \citenamefont {{Di Candia}}, \citenamefont {Deppe}, \citenamefont {Eder},
  \citenamefont {Zhong}, \citenamefont {Ihmig}, \citenamefont {Haeberlein},
  \citenamefont {Baust}, \citenamefont {Hoffmann}, \citenamefont {Ballester},
  \citenamefont {Inomata}, \citenamefont {Yamamoto}, \citenamefont {Nakamura},
  \citenamefont {Solano}, \citenamefont {Marx},\ and\ \citenamefont
  {Gross}}]{Menzel2012a}%
  \BibitemOpen
  \bibfield  {author} {\bibinfo {author} {\bibfnamefont {E.~P.}\ \bibnamefont
  {Menzel}}, \bibinfo {author} {\bibfnamefont {R.}~\bibnamefont {{Di Candia}}},
  \bibinfo {author} {\bibfnamefont {F.}~\bibnamefont {Deppe}}, \bibinfo
  {author} {\bibfnamefont {P.}~\bibnamefont {Eder}}, \bibinfo {author}
  {\bibfnamefont {L.}~\bibnamefont {Zhong}}, \bibinfo {author} {\bibfnamefont
  {M.}~\bibnamefont {Ihmig}}, \bibinfo {author} {\bibfnamefont
  {M.}~\bibnamefont {Haeberlein}}, \bibinfo {author} {\bibfnamefont
  {A.}~\bibnamefont {Baust}}, \bibinfo {author} {\bibfnamefont
  {E.}~\bibnamefont {Hoffmann}}, \bibinfo {author} {\bibfnamefont
  {D.}~\bibnamefont {Ballester}}, \bibinfo {author} {\bibfnamefont
  {K.}~\bibnamefont {Inomata}}, \bibinfo {author} {\bibfnamefont
  {T.}~\bibnamefont {Yamamoto}}, \bibinfo {author} {\bibfnamefont
  {Y.}~\bibnamefont {Nakamura}}, \bibinfo {author} {\bibfnamefont
  {E.}~\bibnamefont {Solano}}, \bibinfo {author} {\bibfnamefont
  {A.}~\bibnamefont {Marx}}, \ and\ \bibinfo {author} {\bibfnamefont
  {R.}~\bibnamefont {Gross}},\ }\href@noop {} {\bibfield  {journal} {\bibinfo
  {journal} {Physical Review Letters}\ }\textbf {\bibinfo {volume} {109}},\
  \bibinfo {pages} {250502} (\bibinfo {year} {2012})}\BibitemShut {NoStop}%
\bibitem [{\citenamefont {Forgues}\ \emph {et~al.}(2014)\citenamefont
  {Forgues}, \citenamefont {Lupien},\ and\ \citenamefont
  {Reulet}}]{Forgues2014}%
  \BibitemOpen
  \bibfield  {author} {\bibinfo {author} {\bibfnamefont {J.-C.}\ \bibnamefont
  {Forgues}}, \bibinfo {author} {\bibfnamefont {C.}~\bibnamefont {Lupien}}, \
  and\ \bibinfo {author} {\bibfnamefont {B.}~\bibnamefont {Reulet}},\
  }\href@noop {} {\bibfield  {journal} {\bibinfo  {journal} {Physical Review
  Letters}\ }\textbf {\bibinfo {volume} {113}},\ \bibinfo {pages} {043602}
  (\bibinfo {year} {2014})}\BibitemShut {NoStop}%
\bibitem [{\citenamefont {Forgues}\ \emph {et~al.}(2015)\citenamefont
  {Forgues}, \citenamefont {Lupien},\ and\ \citenamefont
  {Reulet}}]{Forgues2015}%
  \BibitemOpen
  \bibfield  {author} {\bibinfo {author} {\bibfnamefont {J.-C.}\ \bibnamefont
  {Forgues}}, \bibinfo {author} {\bibfnamefont {C.}~\bibnamefont {Lupien}}, \
  and\ \bibinfo {author} {\bibfnamefont {B.}~\bibnamefont {Reulet}},\
  }\href@noop {} {\bibfield  {journal} {\bibinfo  {journal} {Physical Review
  Letters}\ }\textbf {\bibinfo {volume} {114}},\ \bibinfo {pages} {130403}
  (\bibinfo {year} {2015})}\BibitemShut {NoStop}%
\bibitem [{\citenamefont {da~Silva}\ \emph {et~al.}(2010)\citenamefont
  {da~Silva}, \citenamefont {Bozyigit}, \citenamefont {Wallraff},\ and\
  \citenamefont {Blais}}]{DaSilva2010}%
  \BibitemOpen
  \bibfield  {author} {\bibinfo {author} {\bibfnamefont {M.~P.}\ \bibnamefont
  {da~Silva}}, \bibinfo {author} {\bibfnamefont {D.}~\bibnamefont {Bozyigit}},
  \bibinfo {author} {\bibfnamefont {A.}~\bibnamefont {Wallraff}}, \ and\
  \bibinfo {author} {\bibfnamefont {A.}~\bibnamefont {Blais}},\ }\href@noop {}
  {\bibfield  {journal} {\bibinfo  {journal} {Physical Review A}\ }\textbf
  {\bibinfo {volume} {82}},\ \bibinfo {pages} {043804} (\bibinfo {year}
  {2010})}\BibitemShut {NoStop}%
\bibitem [{\citenamefont {Gabelli}\ \emph {et~al.}(2004)\citenamefont
  {Gabelli}, \citenamefont {Reydellet}, \citenamefont {F\`{e}ve}, \citenamefont
  {Berroir}, \citenamefont {Pla\c{c}ais}, \citenamefont {Roche},\ and\
  \citenamefont {Glattli}}]{Gabelli2004}%
  \BibitemOpen
  \bibfield  {author} {\bibinfo {author} {\bibfnamefont {J.}~\bibnamefont
  {Gabelli}}, \bibinfo {author} {\bibfnamefont {L.-H.}\ \bibnamefont
  {Reydellet}}, \bibinfo {author} {\bibfnamefont {G.}~\bibnamefont {F\`{e}ve}},
  \bibinfo {author} {\bibfnamefont {J.-M.}\ \bibnamefont {Berroir}}, \bibinfo
  {author} {\bibfnamefont {B.}~\bibnamefont {Pla\c{c}ais}}, \bibinfo {author}
  {\bibfnamefont {P.}~\bibnamefont {Roche}}, \ and\ \bibinfo {author}
  {\bibfnamefont {D.~C.}\ \bibnamefont {Glattli}},\ }\href@noop {} {\bibfield
  {journal} {\bibinfo  {journal} {Physical Review Letters}\ }\textbf {\bibinfo
  {volume} {93}},\ \bibinfo {pages} {056801} (\bibinfo {year}
  {2004})}\BibitemShut {NoStop}%
\bibitem [{\citenamefont {Menzel}\ \emph {et~al.}(2010)\citenamefont {Menzel},
  \citenamefont {Deppe}, \citenamefont {Mariantoni}, \citenamefont {{Araque
  Caballero}}, \citenamefont {Baust}, \citenamefont {Niemczyk}, \citenamefont
  {Hoffmann}, \citenamefont {Marx}, \citenamefont {Solano},\ and\ \citenamefont
  {Gross}}]{Menzel2010}%
  \BibitemOpen
  \bibfield  {author} {\bibinfo {author} {\bibfnamefont {E.~P.}\ \bibnamefont
  {Menzel}}, \bibinfo {author} {\bibfnamefont {F.}~\bibnamefont {Deppe}},
  \bibinfo {author} {\bibfnamefont {M.}~\bibnamefont {Mariantoni}}, \bibinfo
  {author} {\bibfnamefont {M.~A.}\ \bibnamefont {{Araque Caballero}}}, \bibinfo
  {author} {\bibfnamefont {A.}~\bibnamefont {Baust}}, \bibinfo {author}
  {\bibfnamefont {T.}~\bibnamefont {Niemczyk}}, \bibinfo {author}
  {\bibfnamefont {E.}~\bibnamefont {Hoffmann}}, \bibinfo {author}
  {\bibfnamefont {A.}~\bibnamefont {Marx}}, \bibinfo {author} {\bibfnamefont
  {E.}~\bibnamefont {Solano}}, \ and\ \bibinfo {author} {\bibfnamefont
  {R.}~\bibnamefont {Gross}},\ }\href@noop {} {\bibfield  {journal} {\bibinfo
  {journal} {Physical Review Letters}\ }\textbf {\bibinfo {volume} {105}},\
  \bibinfo {pages} {100401} (\bibinfo {year} {2010})}\BibitemShut {NoStop}%
\bibitem [{\citenamefont {Zakka-Bajjani}\ \emph {et~al.}(2010)\citenamefont
  {Zakka-Bajjani}, \citenamefont {Dufouleur}, \citenamefont {Coulombel},
  \citenamefont {Roche}, \citenamefont {Glattli},\ and\ \citenamefont
  {Portier}}]{Zakka-Bajjani2010}%
  \BibitemOpen
  \bibfield  {author} {\bibinfo {author} {\bibfnamefont {E.}~\bibnamefont
  {Zakka-Bajjani}}, \bibinfo {author} {\bibfnamefont {J.}~\bibnamefont
  {Dufouleur}}, \bibinfo {author} {\bibfnamefont {N.}~\bibnamefont
  {Coulombel}}, \bibinfo {author} {\bibfnamefont {P.}~\bibnamefont {Roche}},
  \bibinfo {author} {\bibfnamefont {D.~C.}\ \bibnamefont {Glattli}}, \ and\
  \bibinfo {author} {\bibfnamefont {F.}~\bibnamefont {Portier}},\ }\href@noop
  {} {\bibfield  {journal} {\bibinfo  {journal} {Physical Review Letters}\
  }\textbf {\bibinfo {volume} {104}},\ \bibinfo {pages} {206802} (\bibinfo
  {year} {2010})}\BibitemShut {NoStop}%
\bibitem [{\citenamefont {Kenney}\ and\ \citenamefont
  {Keeping}(1951)}]{Kenney1951}%
  \BibitemOpen
  \bibfield  {author} {\bibinfo {author} {\bibfnamefont {J.~F.}\ \bibnamefont
  {Kenney}}\ and\ \bibinfo {author} {\bibfnamefont {E.~S.}\ \bibnamefont
  {Keeping}},\ }\href@noop {} {\emph {\bibinfo {title} {{Mathematics of
  Statistics, Part Two}}}},\ \bibinfo {edition} {2nd}\ ed.\ (\bibinfo
  {publisher} {Van Nostrand, New York},\ \bibinfo {year} {1951})\BibitemShut
  {NoStop}%
\bibitem [{\citenamefont {Cahill}\ and\ \citenamefont
  {Glauber}(1969)}]{Cahill1969a}%
  \BibitemOpen
  \bibfield  {author} {\bibinfo {author} {\bibfnamefont {K.}~\bibnamefont
  {Cahill}}\ and\ \bibinfo {author} {\bibfnamefont {R.~J.}\ \bibnamefont
  {Glauber}},\ }\href@noop {} {\bibfield  {journal} {\bibinfo  {journal}
  {Physical Review}\ }\textbf {\bibinfo {volume} {177}},\ \bibinfo {pages}
  {1857} (\bibinfo {year} {1969})}\BibitemShut {NoStop}%
\bibitem [{\citenamefont {Gardiner}\ and\ \citenamefont
  {Zoller}(2004)}]{Gardiner2004}%
  \BibitemOpen
  \bibfield  {author} {\bibinfo {author} {\bibfnamefont {C.~W.}\ \bibnamefont
  {Gardiner}}\ and\ \bibinfo {author} {\bibfnamefont {P.}~\bibnamefont
  {Zoller}},\ }\href@noop {} {\emph {\bibinfo {title} {{Quantum Noise}}}},\
  \bibinfo {edition} {3rd}\ ed.\ (\bibinfo  {publisher} {Springer Berlin /
  Heidelberg},\ \bibinfo {year} {2004})\BibitemShut {NoStop}%
\bibitem [{\citenamefont {Blanter}\ and\ \citenamefont
  {B\"{u}ttiker}(2000)}]{Blanter2000}%
  \BibitemOpen
  \bibfield  {author} {\bibinfo {author} {\bibfnamefont {Y.}~\bibnamefont
  {Blanter}}\ and\ \bibinfo {author} {\bibfnamefont {M.}~\bibnamefont
  {B\"{u}ttiker}},\ }\href@noop {} {\bibfield  {journal} {\bibinfo  {journal}
  {Physics Reports}\ }\textbf {\bibinfo {volume} {336}},\ \bibinfo {pages} {1}
  (\bibinfo {year} {2000})}\BibitemShut {NoStop}%
\bibitem [{\citenamefont {Loudon}(2000)}]{Loudon2000}%
  \BibitemOpen
  \bibfield  {author} {\bibinfo {author} {\bibfnamefont {R.}~\bibnamefont
  {Loudon}},\ }\href@noop {} {\emph {\bibinfo {title} {{The Quantum Theory of
  Light}}}},\ \bibinfo {edition} {3rd}\ ed.\ (\bibinfo  {publisher} {Oxford
  University Press},\ \bibinfo {year} {2000})\BibitemShut {NoStop}%
\bibitem [{\citenamefont {Widder}(1946)}]{Widder1946}%
  \BibitemOpen
  \bibfield  {author} {\bibinfo {author} {\bibfnamefont {D.~V.}\ \bibnamefont
  {Widder}},\ }\href@noop {} {\emph {\bibinfo {title} {{The Laplace
  Transform}}}}\ (\bibinfo  {publisher} {Princeton University Press},\ \bibinfo
  {year} {1946})\BibitemShut {NoStop}%
\bibitem [{Note1()}]{Note1}%
  \BibitemOpen
  \bibinfo {note} {Alternatively, it is common in quantum optical experiments
  to measure related quantities such as Mandel's $Q$ parameter ($Q=\protect
  \mathcal {F}-1$) or the degree of second-order coherence
  ($g^{(2)}(0)=1+Q/\protect \langle n\protect \rangle $).}\BibitemShut {Stop}%
\bibitem [{\citenamefont {Grosse}\ \emph {et~al.}(2007)\citenamefont {Grosse},
  \citenamefont {Symul}, \citenamefont {Stobińska}, \citenamefont {Ralph},\
  and\ \citenamefont {Lam}}]{Grosse2007}%
  \BibitemOpen
  \bibfield  {author} {\bibinfo {author} {\bibfnamefont {N.~B.}\ \bibnamefont
  {Grosse}}, \bibinfo {author} {\bibfnamefont {T.}~\bibnamefont {Symul}},
  \bibinfo {author} {\bibfnamefont {M.}~\bibnamefont {Stobińska}}, \bibinfo
  {author} {\bibfnamefont {T.~C.}\ \bibnamefont {Ralph}}, \ and\ \bibinfo
  {author} {\bibfnamefont {P.~K.}\ \bibnamefont {Lam}},\ }\href@noop {}
  {\bibfield  {journal} {\bibinfo  {journal} {Physical Review Letters}\
  }\textbf {\bibinfo {volume} {98}},\ \bibinfo {pages} {2} (\bibinfo {year}
  {2007})}\BibitemShut {NoStop}%
\bibitem [{\citenamefont {Lemonde}\ \emph {et~al.}(2014)\citenamefont
  {Lemonde}, \citenamefont {Didier},\ and\ \citenamefont
  {Clerk}}]{Lemonde2014}%
  \BibitemOpen
  \bibfield  {author} {\bibinfo {author} {\bibfnamefont {M.-A.}\ \bibnamefont
  {Lemonde}}, \bibinfo {author} {\bibfnamefont {N.}~\bibnamefont {Didier}}, \
  and\ \bibinfo {author} {\bibfnamefont {A.~A.}\ \bibnamefont {Clerk}},\
  }\href@noop {} {\bibfield  {journal} {\bibinfo  {journal} {Physical Review
  A}\ }\textbf {\bibinfo {volume} {90}},\ \bibinfo {pages} {063824} (\bibinfo
  {year} {2014})}\BibitemShut {NoStop}%
\bibitem [{\citenamefont {Hatridge}\ \emph {et~al.}(2011)\citenamefont
  {Hatridge}, \citenamefont {Vijay}, \citenamefont {Slichter}, \citenamefont
  {Clarke},\ and\ \citenamefont {Siddiqi}}]{Hatridge2011}%
  \BibitemOpen
  \bibfield  {author} {\bibinfo {author} {\bibfnamefont {M.}~\bibnamefont
  {Hatridge}}, \bibinfo {author} {\bibfnamefont {R.}~\bibnamefont {Vijay}},
  \bibinfo {author} {\bibfnamefont {D.~H.}\ \bibnamefont {Slichter}}, \bibinfo
  {author} {\bibfnamefont {J.}~\bibnamefont {Clarke}}, \ and\ \bibinfo {author}
  {\bibfnamefont {I.}~\bibnamefont {Siddiqi}},\ }\href@noop {} {\bibfield
  {journal} {\bibinfo  {journal} {Physical Review B}\ }\textbf {\bibinfo
  {volume} {83}},\ \bibinfo {pages} {134501} (\bibinfo {year}
  {2011})}\BibitemShut {NoStop}%
\end{thebibliography}%

\end{document}